\documentstyle[12pt,aasms4]{article}
\def\etal{{\it et.~al.}}
\def\eg{{\it e.g.}}

\begin{document}

\received{}
\accepted{}

\title{RADIO LOUD AND RADIO QUIET ACTIVE GALACTIC NUCLEI}
\author{\it Chun Xu\altaffilmark{1}, Mario Livio and Stefi Baum} 

\affil{Space Telescope Science Institute\\ 3700 San Martin Drive
\\ Baltimore, MD  21218
\\ chunxu@stsci.edu, mlivio@stsci.edu, sbaum@stsci.edu}

\altaffiltext{1}{Also at: Department of Astronomy, University of Maryland,
College Park, MD 20742}

\begin{abstract}
We generated a sample of 409 AGNs for which both the radio luminosity at
5 GHz and the line luminosity in [OIII] 5007 have been measured. 
The radio luminosity spans a range of ten orders of magnitude,
and the [OIII] line luminosity spans a range of eight orders of magnitude
--- both considerably larger than the ranges in previous studies. We show
that these two quantities are correlated in a similar way for both radio-loud
and radio-quiet AGNs. We demonstrate that the observed correlation can be
explained in terms of a model in which jets are accelerated and collimated
by a vertical magnetic field.
\end{abstract}


\section{INTRODUCTION}

The ``unified'' scheme of active galactic nuclei (AGNs) has been very 
successful in explaining a variety of AGN properties on the basis of the
viewing angle $\theta$ (see \eg, review by Urry \& Padovani 1995). It is
very clear, however, that more physical parameters are required to
construct a truly ``unified'' model of all AGNs (\eg, Blandford 1990).
In particular, it is by now well established that AGNs fall into two
families in terms of their radio power (as measured for example by their
5~GHz luminosities), the ``radio-loud'' and ``radio-quiet'' AGNs (\eg, Baum
\& Heckman 1989; Miller, Rawlings \& Saunders 1993).  Phenomenologically,
radio-louds are always associated with large scale radio jets and lobes
while the radio-quiets have very little or weak radio emitting ejecta.
Radio-loud AGNs are almost always associated with early type galaxies, while the
radio-quiet ones are mostly found in spirals and S0s.

One of the ways to tackle observationally the question of the additional
fundamental parameters (other than the viewing angle $\theta$), is to
examine the AGN luminosities in a variety of wavebands.  In particular,
it has been shown that a correlation exists between the radio luminosity
($L_{5~GHz}$) and the [OIII] 5007 narrow line luminosity ($L_{[OIII]}$) (\eg,
Rawlings \& Saunders 1991; Baum \& Heckman 1989).

In the present work, we study the $L_{5~GHz}$--$L_{[OIII]}$ and related
correlations for a much larger data set, covering a very wide range of
luminosities.  We then use the obtained results in combination with
recent theoretical developments in an attempt to place constraints on
possible scenarios for the nature of radio-loud and radio-quiet AGNs.

In Section~2 we describe the samples used and the data.  The results are
presented in Section~3, and discussed in Section~4. A summary and
conclusions follow.

\section{SAMPLES AND DATA HANDLING}

All the data were compiled from the literature and through the NASA
Extragalactic Database (NED).  Our original sample included four
categories of AGNs: (1)~Radio sources (Bennett 1962; Smith \& Spinrad
1980; Zirbel \& Baum 1995; Condon, Frayer \& Broderick 1991), (2)~Seyfert
galaxies (Lipovetsky, Neizvestung \& Neizvestnara 1988; Dahari \&
Robertis 1988; Whittle 1992), (3)~BL~Lac objects (Veron-Cetty \& Veron
1996; Padovani \& Giommi 1995), and (4)~Quasars (Schmidt \& Green 1983;
Brotherton 1996; Boroson \& Green 1992).  To these we added many sources
found individually in the literature, thus generating a sample of about
2,000~AGNs. From this sample we selected all the objects for which
measurements of both the radio luminosity at 5~GHz and the luminosity in
the [OIII] 5007 forbidden line existed. We have thus generated a sample
of 409~sources, including: 162~Seyfert galaxies, 136~quasars, 107~radio
galaxies and 4~BL~Lac objects.

Table 1 gives a list of all the sources used, with the relevant
information for each object.  Specifically, we list in columns 1--6 
respectively the
object's IAU name, its catalogue name, its identification 
(a galaxy (G) or a quasar (Q)), the nature of its activity: Seyfert
or Radio Galaxy, its morphology type and its redshift. In columns
7--10 we list respectively: its luminosity in the [OIII] 5007 line,  the 
total radio power at 5~GHz, the core radio power at 5~GHz, the
x-ray luminosity in the 2--10~keV band and the corresponding references.  
In the last column (column 15) we remark if the object is regarded as
radio-loud (L) or radio-quiet (Q) in our study. A more detailed notation is 
attached at the end of Table 1. Almost all the data presented on
the luminosities represent actual measurements, with very few (\eg, core
radio powers from Zirbel \& Baum 1995) upper and lower limits.  As a
rule, if a certain quantity was found to have several different quoted
values, the one closest to the mean was taken.

The radio powers at 5~GHz given in Table~1 were either taken directly
from the literature, or calculated from the given fluxes.  A value of
the Hubble constant of H$_0=50$~km~s$^{-1}$ Mpc$^{-1}$ has been assumed
throughout.  In calculating the radio power of sources for which the
spectral index was not available (assuming a power law
$S_{\nu}\,\propto\,\nu^{-\alpha}$), a mean index of 0.75 was adopted.

The luminosities in the [OIII] 5007 line were mostly calculated from the
fluxes found in the literature (without reddening corrections, since
very few of the latter are available, \eg, Koski 1978). In a case in
which only the combined fluxes of [OIII]~5007 and 4959 were given
(Steiner 1981), the flux in the 5007 line was taken to be 3/4 of the
combined flux.  In two cases in which only equivalent widths of
[OIII]~5007 were given (Brotherton 1996; Boroson \& Green 1992), the
continua were determined from the corresponding spectra.

The x-ray luminosities in the 2--10~keV band were calculated from the
integrated fluxes.  In a case in which only the HEAO/A-2 count rates was
given (Ceca \etal~1990), the fluxes were calculated assuming a mean
energy index of 0.65.

The redshifts were obtained from the literature and cross checked
through the NED.

While clearly the use of many sources for the data makes our sample
inhomogeneous in terms of the errors involved, this has a very little
effect on our conclusions, since the data now span 8~orders of
magnitudes in the [OIII] luminosity and 10~orders of magnitude in the
radio luminosity.

\section{ANALYSIS AND RESULTS}

In Fig.~1, we present the radio luminosity at 5~GHz, $L_{5~GHz}$, as a
function of the [OIII]~5007 line luminosity, $L_{[OIII]}$, for all the
sources in our sample.  The separation into the two families of 
radio-loud and radio-quiet AGNs with a significant gap between them is
immediately apparent.  The radio luminosities are different by a factor
$10^3$--$10^4$ between the two groups (at a given [OIII] luminosity).
We do find a small group ($\sim3$\% of the sample) of objects which
appear to occupy the region between the two main families.
These tentative ``intermediate'' objects have been represented by filled
circles in Fig.~1.

Linear fits to the radio-loud and radio-quiet AGNs (excluding the
intermediate objects) give:
\begin{eqnarray}
\log\left(L_{5~GHz}\right) &=
&(0.61\pm0.07)~\log\left(L_{[OIII]}\right)+(3.7\pm0.6)~~{\rm radio~loud}
\nonumber\\
\log\left(L_{5~GHz}\right) &=
&(0.45\pm0.07)~\log\left(L_{[OIII]}\right)+(5.6\pm0.6)~~{\rm radio~quiet}
\end{eqnarray}
Here, $L_{5~GHz}$ is the radio luminosity in Watts Hz$^{-1}$ Sr$^{-1}$,
and $L_{[OIII]}$ is the [OIII] line luminosity in Watts.
In determining whether an object belongs to the radio-loud, radio-quiet or
intermediate group, we first selected out two distinct groups of objects
(namely radio-louds and radio-quiets) based on the distribution
in the $L_{5 GHz}$--$L_{[OIII]}$ diagram (Fig.~1).  We then found 
linear fits to the two groups respectively. 
We considered  the objects whose radio luminosities (at 5 GHz) are
higher than the ``radio-loud" fitting minus 2.75 $\sigma$ as radio-loud,
those whose radio luminosities are less than the ``radio-quiet" fitting
plus 2.5 $\sigma$ as radio-quiet, and those which lie between the two 
groups as intermediate. This way of defining radio-loud and radio-quiet 
is to some extent semi-empirical. The reason we chose slightly different 
criteria in determining the
radio-loud and radio-quiet membership was to achieve the best visual
representation  of the two groups in Fig.~1.

We note that the samples used by previous authors (\eg, radio quiet
quasars by Miller \etal~1993, radio-loud objects by Rawlings 1994)
represent sub-groups of our sample.  These authors found a relation with
a somewhat steeper slope ($\sim0.85$--1.0) for the subgroups.

In Fig.~2a, we present the same data, now indicating the different
classes of AGNs (note that a few of the objects might have been
misclassified in the literature).  What becomes immediately clear from
this figure is that by plotting $L_{5~GHz}$ vs.\ $L_{[OIII]}$ only for
some {\it individual classes\/} of AGNs (\eg, quasars), one could obtain
a distribution of points that would have looked like a relation with a
rather different slope (see Fig.~2b).  In such a case it would have been
difficult to determine whether this indeed represents a different
functional dependence of $L_{5~GHz}$ or $L_{[OIII]}$, or whether this is
merely an artifact of examining, for example, only the rightmost edges
of two separate linear relations (as in the case of quasars, Figs.~2a, 2b).
For the moment we will assume that the latter interpretation is correct,
since our basic assumption is one of an underlying ``unified'' scheme.
We do note however that extreme caution should be exercised in attempts to
determine the properties of subclasses of AGNs, and that in principle,
the former interpretation may be valid (\eg, that radio-loud and radio-quiet
QSOs are more fundamentally associated).

In order to examine the question of whether the distinction between the
radio-loud and radio-quiet families is mainly a result of the extended
radio emission, we plotted in Fig.~3 the 5~GHz core radio luminosity
against the [OIII] line luminosity.  As we can see from the figure, the
gap between the radio-loud and radio-quiet groups is much less
pronounced in this case (although instrumental effects introduce
uncertainties; see also Nelson \& Whittle 1996 and Sadler \etal~1995), 
but the two groups are still discernible.
Interestingly, most of the ``intermediate'' sources of Fig.~1 become
essentially indistinguishable from the radio-loud sources in Fig.~3.
This may indicate a close relation between the intermediate sources and
the radio louds.  A linear fit to the core radio luminosity of the
radio-louds gives
\begin{equation}
\log\left(L_{core}\right)=(0.78\pm0.07)~\log\left(L_{[OIII]}\right)
-(3.3\pm1.1)~~,
\end{equation}
where $L_{core}$ is the core radio luminosity (at 5~GHz) in 
Watts~Hz$^{-1}$ Sr$^{-1}$.

In order to further clarify the properties of the [OIII] emission, we plot
in Fig.~4 the x-ray luminosity in the 2--10~keV band as a function of
the [OIII] luminosity.  As can be seen from the figure, a clear positive
correlation exists, and a linear fit gives 
\begin{equation}
\log\left(L_{HX}\right)=(1.01\pm0.08)~\log(L_{[OIII]})+(8.6\pm0.7)~~,
\end{equation}
where $L_{HX}$ is the x-ray luminosity (in the 2--10~keV range) in
erg~s$^{-1}$.  Individual fits to the radio-loud (plus intermediate) sample
and the radio-quiet sample give slopes of $0.89\pm0.19$ and $0.95\pm0.10$ 
respectively.

It is important to note right away that the radio-louds and radio-quiets
share the same $L_{HX}$--$L_{[OIII]}$ relation, implying probably that the same
physical process relates these quantities in the two families (we will
return to this point later).

In order to examine the potential effects of the environment, we also
plotted the data of Fig.~1 distinguishing among the different types of
host galaxies (in cases in which the latter have been identified).  This
is presented in Fig.~5.  This figure confirms that {\it all\/} the radio
louds are elliptical or S0~galaxies, while almost all of the spirals are
radio quiet.  It is interesting to note though that some of the
``intermediate'' objects are spirals. We should note, however, that quite 
a few of the morphologies are ambiguous (in particular the distinction
between ellipticals and S0s) and therefore caution should be
exercised in attempts to draw conclusions on the basis of morphology.

\section{DISCUSSION}

We have found that over a very wide range in luminosities, 
AGNs separate into two classes of objects, radio-loud
and radio-quiet, and that the two classes obey two parallel 
relations between their radio and
[OIII] luminosities, of the form
\begin{equation}
L_{5~GHz}\sim L^{\beta}_{[OIII]}~~,
\end{equation}
with $\beta\sim0.5$ (Fig.~1 and eq.~1).  
At this point we need to consider whether the observed relationships
are true correlations or whether selection effects can be dominating
what we are seeing. First, we need to ask, are there really 
two distinct classes of radio-quiet and radio-loud types 
of AGN or is the apparent gap region merely
an artifact of the selection process? In particular, many, though not
all, of the radio loud objects come from radio flux density selected
samples and many, though not all, of the radio quiet objects come from
optically selected samples.
 
We do not think that selection effects are responsible for the apparent
population of the $L_{radio}-L_{[OIII]}$ plane for the following reasons: (i)
radio flux density selected samples always include objects of low radio
luminosity at low redshift and of progressively higher and higher radio
luminosity with redshift. For example, the 3CR cut off in radio
flux  is about 9 Jy at 178 MHz (Bennett 1962); if we take the spectral
index as 0.75 (which is a typical value for AGNs), then the extrapolated 
radio flux at 5 GHz is 0.74 Jy. This corresponds to $L_{5GHz} \sim 2.3 \times
10^{21} W/Hz/Sr$ at z=0.003 and $L_{5GHz} \sim 2.56 \times 10^{22} W/Hz/Sr$
at z=0.01. These values lie within the radio-quiet group in our 
classification (Fig.~1). Actually, we do find sources from the 3CR
catalogue which are indeed radio-quiets (3C71 for example).  
Likewise, purely optically selected samples of AGN 
(e.g., Padovani 1993; Miller \etal~1990 and Kellermann \etal~1989)
also show the radio-quiet-radio-loud dichotomy, with roughly 1 in 10 of
optically selected quasars falling into the radio loud category.
(ii) Perhaps more importantly, the apparent dichotomy in
properties we have found  
is not a dichotomy of either $L_{radio}$ or $L_{[OIII]}$, 
but the {\it ratio} of these 
two quantities. It is clear that selection effects can cause
us to find sources with high, on average, radio luminosity when we select radio
flux density limited samples; however, there is no a priori reason why
such samples should show a limited and specific ratio of radio to line
luminosity. The absence of identified sources with
high radio luminosity and low line luminosity in radio flux density
selected samples has been known for years (e.g., Baum and Heckman,
1989); sources with high radio luminosity invariably have high accompanying
line luminosities.
The real surprise is then that the {\it converse} does 
not also appear to be true.
That is, if one selects AGN of high line luminosity at any given redshift, one
finds two distinct classes of objects; those with high radio 
luminosity and those without. Said in a different way, there are sources with
very high line luminosity which do not have accompanying very high 
radio luminosity.  This is the radio-quiet radio-loud
dichotomy that has been known for many years (e.g., Antonucci 1993 
and references therein);
and the fundamental paradox: the nuclei of active
galaxies can produce copious amounts of line luminosity (and UV through
X-ray luminosity) without producing large amounts of radio luminosity,
however the nuclei of active galaxies cannot produce large amounts of
radio luminosity without producing concomitant large amounts of
line and UV/X-ray luminosity. 

\subsection {The Physical Origin Of The $L_{radio} \-- L_{[OIII]}$ Relation}

We will now attempt to
understand the origin of this relation in terms of the physical
processes involved.

There exists strong observational evidence that suggests that $L_{[OIII]}$
is one of the best orientation independent measures of the intrinsic 
luminosity of the nuclei
of AGNs (\eg, Miller \etal~1992; Jackson \& Browne 1991; Mulchaey \etal~
1994).  This fact, in combination with the data presented in Fig.~4,
suggests that $L_{[OIII]}$ is proportional to the accretion rate through
the accretion disk, $\dot{M}_{acc}$.  We will therefore assume that 
$L_{[OIII]}$ is proportional to $\dot{M}_{acc}$,
$L_{[OIII]}\propto\dot{M}_{acc}$.

We will now attempt to obtain a general relation between the accretion
rate through the disk and the mass flux into the jet, $\dot{M}_j$ (see
\eg, Pringle 1993; Tout \& Pringle 1996; Livio 1997).  

To this goal, we first note that the most promising models for jet
acceleration and collimation involve an accretion disk that is threaded
by a large scale, vertical magnetic field (\eg, Blandford \& Payne 1982;
K\"onigl 1989; and see Livio 1997 for a review).  We will now make the
following simple assumptions: (i)~The accretion disk is largely a standard
(geometrically thin),
Shakura-Sunyaev (1973) disk.  (ii)~The jet velocity, $V_j$, is of the order of
the Keplerian velocity in the inner disk. (iii)~The vertical magnetic
field component is of the order of the azimuthal one, $B_z\sim B_{\phi}$.
A detailed justification of these assumptions can be found in Livio
(1999) and references therein.  Realizing that the back
pressure from the jet on the disk is given by
$P_{jet}\sim\dot{M}_j~V_j/R^2$, where $R$ is the radius from which the
jet originates, and using assumptions (i)--(iii) above, we obtain
\begin{equation}
{{B_z^2/8\pi}\over{P_g}}\sim
{{\dot{M}_j}\over{\dot{M}_{acc}}}~
{H\over R}\alpha~~.
\end{equation}
Here $P_g$ is the gas pressure, $H$ is the disk half-thickness and
$\alpha$ is the Shakura-Sunyaev (1973) viscosity parameter.  If we
assume in addition that the disk viscosity is generated by a dynamo,
which in turn is powered by MHD turbulence (\eg, Hawley, Gammie \& Balbus
1995; Stone \etal~1996; Brandenburg \etal~1995), then
$\alpha\sim B_D^2/(4\pi~P_g)$, where $B_D$ is the magnetic field in the
disk.  Substituting this into eq.~(5) gives
\begin{equation}
{{B_z}\over{B_D}}\sim
\left[
{{\dot{M}_j}\over{\dot{M}_{acc}}}~{{H}\over{R}}
\right]^{1/2}~~.
\end{equation}

An independent relation between $B_z$ and $B_D$ can be obtained if we
make an assumption about the origin of the large-scale vertical field.
In principle, such a field can either be advected inwards by the
accreting matter (\eg, Blandford \& Payne 1982; K\"onigl 1989; Pelletier
\& Pudritz 1992), or it can be generated locally by the same dynamo
processes which generate the disk viscosity (Tout \& Pringle 1996).  If
we assume the latter to be true, then the large scale field may be
obtained through the reconnections of magnetic loops (leading to an
inverse cascade process) which have a length distribution of the form
$n(\ell)\sim\ell^{-\delta}$.  
In such a case, it can be easily
shown that (Tout \& Pringle 1996; Livio 1997)
\begin{equation}
{{B_z}\over{B_D}}\sim
\left({{H}\over{R}}\right)^{\delta-1}~~.
\end{equation}

Combining eqs.~(6) and (7) we obtain
\begin{equation}
{{\dot{M}_j}\over{\dot{M}_{acc}}}
\sim
\left({{H}\over{R}}\right)^{2\delta-3}~~.
\end{equation}

Observations of accretion disks, jets and outflows in young stellar
objects, supersoft x-ray sources and
cataclysmic variables and theoretical models suggest that $\delta$ is in
the range 1.7--3.4 (Livio 1997; Tout \& Pringle 1996). Therefore,
assuming that the jet formation mechanism is similar in all the classes
of objects which produce jets, and noting that H/R is approximately 
constant in standard disks (Shakura \& Sunyaev 1973),
we find that $\dot{M}_j$ is roughly
proportional to $\dot{M}_{acc}$, $\dot{M}_j\,\propto\,\dot{M}_{acc}$.

 The final ingredient that is needed to explain the relation obtained in
Fig.~1 is a relation between $L_{5~GHz}$ and $\dot{M}_j$. Since we have 
shown that $L_{[OIII]} \propto \dot{M}_{acc}$, and that $\dot{M}_j \propto
\dot{M}_{acc}$, it is clear that the dependence observed in Fig.~1 
(and in Rawlings 1994) would be obtained if $L_{5~GHz} \propto 
{\dot{M}_j}^\beta$, with $\beta \sim$ 0.5 - 1.0. Observations of individual 
Galactic jets in systems containing black hole accretors 
(e.g., Hjellming \& Rupen 1995; Mirabel \& Rodriguez 1994;
Tavani \etal~1996) indeed suggest that the radio luminosity is proportional
to some power (of order unity) of the mass flow rate into the jet. 
Simple models of radio emission from jets also predict radio luminosities
which are roughly proportional to the mass flux into the jet 
(with the constant of proportionality depending on some power of the magnetic
field strength and on the age of the source, \eg, Bicknell, Dopita \&
O'Dea 1997).
We therefore conclude that the general correlation in Fig.~1 is entirely
consistent with a model in which jets are formed by accretion disks
(around supermassive black holes) which are threaded by a vertical magnetic
field (e.g., Blandford \& Payne 1982; K\"onigl 1989; Ostriker 1997;
Matsumoto \etal~1996; and see Livio 1997 for a review).

A question that needs to be asked at this point is:
can there be important selection effects which are skewing the
slope of the $L_{5GHz}-L_{[OIII]}$ correlation to be less than unity?
This could occur, most naturally, if we were missing a class of 
high radio luminosity, high [OIII] luminosity objects or if we
have systematically underestimated the line luminosity of the high
luminosity objects. However, there is no evidence that this is the
case.  A fit to the slope for only z$\ge$0.2 radio-loud 
sources indicates, if anything, a slightly flatter slope (0.37$\pm$0.12)
than that found either at low (z$\le$0.2) redshifts or taking the sample as
a whole  (0.61$\pm$0.07), though the differences are not statistically
significant. Similarly, the slope of the radio-quiet class is more dominated by
low redshift objects, but shows no evidence for a change for redshifts
less than (slope 0.44$\pm$0.07) or greater than 0.2 (slope 0.45$\pm$0.55).

 It is important to note that the fact that the $L_{5~GHz} - L_{[OIII]}$
relation {\it has almost the same slope} for both the radio-quiet 
and radio-loud AGNs probably indicates that the jet formation mechanism 
is the same in both of these subclasses.

\subsection {The Black Hole Mass}

 Another consequence of Fig.~1 that should be pointed out is the following
(see also Livio 1997). The mass of the central black hole determines the
Eddington luminosity, and therefore the maximum accretion rate which the
system can sustain (this translates into: how far to the right, 
in Fig.~1, the system can be
found). Hence, we can expect that the AGNs containing the most massive 
black holes will occupy the upper right corner of the distribution for
each subclass (radio-loud and radio-quiet).  Interestingly, we find 
that the distribution of the radio-louds extends to somewhat
larger values of $L_{[OIII]}$ (larger $\dot{M}_{acc}$). In order
to further examine the implications of this fact, we show in Fig.~6
the distribution of the sources with respect to their redshifts. We
find that the sources at low redshifts (z $\leq$ 0.2) exhibit the same
range in $\dot{M}_{acc}~(L_{[OIII]})$ in both radio-louds and radio-quiets,
but that the radio-loud sources at higher redshifts extend to higher
values of $\dot{M}_{acc}$. Therefore, if the Eddington luminosity is
indeed the limiting factor, then this finding can be regarded as suggestive 
that the maximum mass of the black holes found in radio-louds is higher
than that in radio-quiets. This result would be consistent with the
fact that the measured black hole masses appear to correlate with
the bulge luminosities (Kormendy \& Richstone 1995). 

\subsection { The Distinction Between Radio-Louds And Radio-Quiets}
 
 A more difficult and long standing question is what distinguishes
the upper group (radio-louds) from the lower one (radio-quiets). Recent 
discussions of this problem can be found, for example, in Blandford
\& Levinson (1995), Fabian \& Rees (1995), Wilson (1996) and Livio (1997).

 Generally, explanations for the existence of these two classes fall into
two different categories: (i) the ones that assume that the central
engines in radio-louds and radio-quiets are the same, but that either the
formation or the propagation of powerful jets is somehow prohibited
in radio-quiets by some external circumstances. (ii) Ones in which it
is assumed that only the central engines of the radio-louds can produce
truly {\it powerful} jets.

In recent work, Livio (1997, 1999) examined the formation of jets in {\it all}
the classes of astrophysical objects which are observed to produce jets.
On the basis of the 
assumption that the jet formation mechanism is {\it the same} in all
the classes of objects, Livio has shown that the following 
{\it conjecture}
is consistent with all the available observational data: the formation of 
{\it powerful} jets requires in addition to an accretion disk threaded by
a vertical field, an additional energy/wind source like a corona or
a source associated with the central
object. More recently, Ogilvie \& Livio (1998) solved for the local
vertical structure of an accretion disk threaded by a poloidal 
magnetic field. By analyzing the dynamics of the transonic outflow in the
disk corona, they showed that a certain potential difference must be overcome
even when the inclination angle between the magnetic field and the 
vertical to the disk surface is larger than 30\arcdeg. Thus, the launching
of an outflow from an accretion disk indeed requires a hot corona
or access to an additional source of energy, in accordance with Livio's
above conjecture. Livio went on to attempt to identify the extra 
energy/wind source
for every jet-producing class. In the case of black hole accretors, 
the general impression has been that this source may be the
black hole spin (since rotational energy can be extracted, e.g., by the 
Blandford \& Znajek (1977) mechanism). 
Observations suggesting that the spins of the
two jet-producing Galactic black holes (GRS 1915+105 and GRO J1655-40)
are high ($a_*$=0.998 and 0.93 respectively, where $a_*$ is the dimensionless
specific angular momentum), while those of other stellar-mass black holes 
(which do not have jets) are very low ($a_* \sim$ 0; Zhang, Cui \& Chen 1997)
seemed consistent with this impression (although the spin determinations
are rather uncertain). However, more recently, 
Ghosh \& Abramowicz (1997), Livio, Ogilvie \& Pringle (1999), and Li (1999)
have shown that the electromagnetic output from the inner disk is
generally expected to dominate over that from the hole. Consequently,
the spin of the hole may not be the ``extra'' energy source in Livio's
conjecture. Rather, the role of the ``wind" from the central source in Livio's
conjecture may be played by gas pressure of the hot atmosphere in ellipticals,
as suggested by Fabian \& Rees (1995). The latter possibility may be supported
by the fact that Fig.~5 shows that the high $L_{5~GHz}$ group contains
quite a few S0 galaxies (but no spirals), in which the central environments
are generally similar to those in ellipticals.

\section{SUMMARY AND CONCLUSIONS}
On the basis of the data collected in the present work and the discussion
in \S 4, we can draw the following (tentative) conclusions:
(1) Both radio-quiet and radio-loud AGNs obey a linear 
$log~L_{5~GHz}$--$log~L_{[OIII]}$ relation, with a nearly identical slope
(but with a shift towards higher radio power for the radio-louds, by a 
factor  $\sim10^3$--$10^4$; see also Rawlings 1994). 
(2) The radio-louds and radio-quiets
share the {\it same} linear correlation between $log~L_x$ and $log~L_{[OIII]}$
(where $L_x$ is the X-ray luminosity in the 2-10 keV range).
(3) Consistently with previous studies, we find that radio-loud AGNs
are found only in elliptical and S0 galaxies (although the distinction
between S0 and elliptical is often ambiguous), while radio-quiets
are mostly spirals and S0s. (4) The observationally determined 
$L_{5~GHz}$--$L_{[OIII]}$ correlation is consistent with a model in which the
radio-emitting jets are formed by an accretion disk which is threaded
by a vertical magnetic field. (5) It is still not entirely clear whether
the distinction between radio-louds and radio-quiets is a consequence of
differences in the central engines of these two classes or whether it
merely reflects differences in the environments.

Acknowledgments: ML acknowledges support from NASA Grant NAG5-6857. This
work has been supported in part by the Director's Discretionary Research
Fund at STScI. We acknowledge useful discussions with Andrew Wilson.

\newpage  
\centerline{\bf Figure Captions}
\oddsidemargin  0.2in
\parindent -0.4in
 
{\bf Fig~1.} The total 5 GHz luminosity {\it vs.} the [OIII] 5007 line
luminosity for our sample of AGNs. The filled circles represent
the intermediate sources (see text).
 
{\bf Fig~2a.} Same as Fig 1., with the types of the AGNs indicated. 

{\bf Fig~2b.} Same as Fig 1., but only the Quasars are presented. 
 
{\bf Fig~3.} The core 5 GHz luminosity {\it vs.} the [OIII] 5007 luminosity.
Most of the objects persented here are radio-louds.
 
{\bf Fig~4.} The hard X-ray luminosity ( 2 -- 10 keV) {\it vs.} the 
[OIII] 5007 luminosity.
 
{\bf Fig~5.} Same as Fig 1., with the morphological type of the host 
galaxies of the AGNs indicated. (Note that the symbols that appear as
filled squares are actually filled circles reside in open squares, in case
in which an ambiguity exists)

{\bf Fig~6.} Same as Fig 1., with the redshift ranges indicated.  

\end{document}